\documentclass[%
 reprint, nofootinbib,
 amsmath,amssymb,
 aps,
]{revtex4-1}

\usepackage{graphicx}
\usepackage{dcolumn}
\usepackage{bm}
\usepackage{hyperref}
\usepackage{braket}
\usepackage{color}
\usepackage{here}

\begin{document}

\title{Neutrino Diffusion within Dark Matter Spikes}

\author{Motoko Fujiwara\textsuperscript{1,2}, Gonzalo Herrera\textsuperscript{3}, Shunsaku Horiuchi\textsuperscript{3,4}}
\affiliation{
\textsuperscript{1}Physik-Department, Technische Universit\"at M\"unchen, James-Franck-Stra\ss{}e, 85748 Garching, Germany,
\\ 
\textsuperscript{2}
Department of Physics, University of Toyama,
3190 Gofuku, Toyama 930-8555, Japan,
\\
\textsuperscript{3} Center for Neutrino Physics, Department of Physics, Virginia Tech, Blacksburg, VA 24061, USA, and
\\
\textsuperscript{4} 
Kavli IPMU (WPI), UTIAS, The University of Tokyo, Kashiwa, Chiba 277-8583, Japan
}

\date{\today}

\begin{abstract}
Multi-messenger observations of astrophysical transients provide powerful probes of the underlying physics of the source as well as beyond the Standard Model effects. We explore transients that can occur in the vicinity of supermassive black holes at the center of galaxies, including tidal disruption events (TDEs), certain types of blazars, or even supernovae. In such environments, the dark matter (DM) density can be extremely high, resembling a dense spike or core. We study a novel effect of neutrino diffusion sustained via frequent scatterings off DM particles in these regions. We show that for transients occurring within DM spikes or cores, the DM-neutrino scattering can delay the arrival of neutrinos with respect to photons, but this also comes with a suppression of the neutrino flux and energy loss. We apply these effects to the specific example of TDEs, and demonstrate that currently unconstrained parameter space of DM-neutrino interactions can account for the sizable $\mathcal{O}$(days) delay of the tentative high-energy neutrinos observed from some TDEs.
\end{abstract}

\maketitle

\section{\label{sec:introduction}
Introduction}

The non-observation of a particle dark matter (DM) signal, from neither direct detection experiments nor electromagnetic and cosmic-ray observations, constraints any particle DM to not couple predominantly to charged particles. It is conceivable, however, that the DM couples predominantly to neutrinos, a possibility which is still only loosely constrained in wide regions of parameter space, in particular where thermal and non-thermal DM production scenarios have not yet been fully tested \cite{Ma:2006km,Lindner:2010rr,Olivares-DelCampo:2017feq, Arguelles:2019ouk, Cardenas:2024ojd}.

Recently, several works have discussed how DM-neutrino scatterings in astrophysical environments can provide world-leading constraints on models where the scattering cross section rises with the incoming energy of the neutrino, \textit{e.g.,} \cite{Arguelles:2017atb,Kelly:2018tyg,Alvey:2019jzx,Murase:2019xqi,Choi:2019ixb,Ferrer:2022kei, Cline:2022qld, Fujiwara:2023lsv, Cline:2023tkp,Carpio:2022sml, Heston:2024ljf, Lin:2024vzy, Ghosh:2021vkt, Lin:2022dbl, DeRomeri:2023ytt}. 
For example, strong astrophysical constraints arise from the non-significant attenuation of the emitted flux of high-energy neutrinos due to scatterings in the DM spikes at the blazar TXS 0506+056 \cite{Ferrer:2022kei, Cline:2022qld}, as well as the non-jetted AGN NGC 1068 \cite{Cline:2023tkp}.
Also, potential constraints have been discussed using tidal disruption events (TDEs) where stars are tidally stripped apart by massive black holes in the centers of galaxies \cite{Fujiwara:2023lsv}. The strength of those constraints stems in large part from the high density of DM particles expected in the vicinity of the central super-massive black hole (SMBH) where the neutrino emission is expected to occur \cite{Murase:2019vdl, Kheirandish:2021wkm, Eichmann:2022lxh}.
On the other hand, in models with constant cross sections or cross sections enhanced at low energies, cosmological probes are typically stronger \cite{Mangano:2006mp,Wilkinson:2014ksa,Escudero:2018thh,Akita:2023yga, Brax:2023tvn}. 

In general, astrophysical systems have driven constraints from the consideration that the initial fluxes of neutrinos shall not be significantly attenuated due to DM-neutrino scatterings. The attenuation of the fluxes can be described by a cascade equation \cite{Arguelles:2017atb, Vincent:2017svp, Cline:2022qld, Fujiwara:2023lsv, HerreraMoreno:2023lvm}. Here, an exponential attenuation term usually dominates over the secondary term accounting for the redistribution of neutrino energies, especially when the initial neutrino flux follows a falling power-law with neutrino energy.

It is possible, however, that for sufficiently large values of the DM-neutrino scattering cross section, and/or coupled with the large DM densities that can be realized in spikes or cores around black holes, the neutrinos enter a diffusion regime. 
For example, if the DM particles self-annihilate, the DM density spike obtains a cored-like center with quasi-constant density, where neutrinos may scatter multiple times before escaping. 
In this case, neutrinos escape over the diffusion timescale, which affects the arrival and spread in time of neutrinos. 
While delays of astrophysical neutrinos induced by scatterings off DM particles have been considered in the intergalactic medium and the Milky Way in previous studies \cite{Murase:2019xqi, Carpio:2022sml, Carpio:2022lqk, Eskenasy:2022aup, Visinelli:2024wyw, KA:2023dyz}, here we estimate for the first time the delays induced by DM at the  \textit{source} of neutrinos. We then correlate the neutrino diffusion time with the suppression of the initial emitted neutrino fluxes.

Our scenario is of particular interest to TDEs, which occur in the vicinity of SMBH at the centers of galaxies where the DM density could be high, and where there are hints of neutrino detections from a sample of TDEs with a consistent delay of ${\cal  O}(100)$ days with respect to the peak of the TDE's electromagnetic signal \cite{Stein:2020xhk, Reusch:2021ztx, Winter:2022fpf, Yuan:2024foi}. Such a delay, although possible in complex scenarios, is not easy to explain with simple astrophysical emission models \cite{Winter_2023}. Here, we explore how DM-neutrino scatterings may contribute significantly to this delay for DM parameters that are currently unconstrained. 
Although TDEs are perhaps the most interesting application, we emphasize that our scenario is generally applicable to neutrino sources embedded in regions where the DM density is high enough for multiple DM-neutrino interactions to occur, \textit{e.g.,} certain blazars or supernovae near the centers of galaxies. 

The paper is organized as follows: In Sec.~\ref{sec:DM_diffusion}, we discuss the general formalism used throughout the paper to quantify the size and density of DM spikes and cores. In addition, we discuss the conditions that must be fulfilled for the system to enter the diffusion regime, and explore impacts on neutrino signatures. In Sec.~\ref{sec:applications}, we apply the formalism to specific neutrino sources such as TDE candidates, from which a delay in neutrino emission has been hinted. Here, we also discuss potential application to future neutrino sources. Lastly, in Sec.~\ref{sec:conclusions} we present our conclusions.

\section{\label{sec:DM_diffusion}
Neutrino diffusion in high DM density cores around SMBHs}

We focus on situations when DM-neutrino scatterings cause neutrinos to diffusively escape dense DM environment, such as those in the vicinity of SMBHs where DM spikes may form.  
A well-known realization of neutrino diffusion is in the cores of collapsed massive stars (\textit{e.g.,} \cite{Janka:2017vlw}). 
Stars with masses $\gtrsim  8 M_\odot$ evolve to form iron cores, which undergo core collapse when electron captures reduce pressure support. 
Subsequently, neutrinos of MeV energies of all flavors are produced in copious numbers through multiple channels. 
During this core-collapse process, the matter density is rapidly increased, the neutrino opacity rises, and the mean free path of neutrinos becomes smaller than the core size. 
As a result, even though the gravitational collapse proceeds on a free-fall time scale of milliseconds, the neutrino emission is set by the neutrino diffusion timescale of several seconds. 
This overall picture was confirmed \cite{Horiuchi:2018ofe} by the observation of neutrinos from SN 1987A arriving over $\sim 10$ seconds  \cite{Kamiokande-II:1987idp,Bionta:1987qt}. 

We first review the DM density distribution around SMBH in Sec.~\ref{sec:spike}, followed by a discussion of when the diffusion limit applies in Sec.~\ref{sec:constraints}. Then, we explore the impacts of the diffusion regime on neutrino signatures in Sec.~\ref{sec:delay}.

\subsection{DM density spikes}\label{sec:spike}

The DM density can be extremely high near BH environments due to gravitational effects, in particular for distances closer than $r  \lesssim  1~\mathrm{pc}$ from the SMBH.
In particular, this occurs when the SMBH grows adiabatically. The adiabatic growth is realized when the dynamical timescale of the accreted particles is much smaller than the growth timescale of the black hole. The accretion timescale of black holes can be estimated with the Sapeter timescale, $t_{\rm 
 S}=M_{\rm{BH}} / \dot{M}_{\rm{Edd}} \simeq 4.5 \times 10^7 \rm{yr}$, while the dynamical timescale can be estimated as $t_{\rm{dyn}}=G M_{\rm{BH}} / \sigma^3$, with $\sigma$ the velocity dispersion of the particles. This condition is generically satisfied for SMBH with masses below $M_{\rm BH} \lesssim 10^{10} M_{\odot}$, see, \textit{e.g.,} \cite{Sigurdsson:2003wu, Herrera:2023nww}.

Under adiabatic growth, the DM density is predicted to form a \textit{spike} where the radial profile follows a power law that is steeper than the original DM profile~\cite{Quinlan:1994ed,Gondolo_1999}.
By matching the spike power law with the original Navarro-Frenk-White (NFW) profile \cite{Navarro:1996gj}, which has $\rho_{\rm  DM}  \propto  r^{-1}$ dependence around the central region, the modified power law is obtained as $\rho_{\rm  DM}  \propto  r^{-  \gamma_{\rm  sp}}$ where $\gamma_{\rm sp}  =  (9  -  2  \gamma)/(4-\gamma)  =  2.33 \cdots$. 

Various effects have been explored that alter the idealized scenario described above. For example, the spike power law can differ depending on the formation history of the SMBH, the merger rate of the Galaxy, and its baryon content and activity \cite{Ullio:2001fb,Merritt:2002vj, Merritt:2006mt}. Also, DM spikes may be disrupted over the course of evolution, \textit{e.g.}, if the DM distribution is gravitationally affected by stars, the spike power law can be shallower, $\gamma_{\rm  sp}  =  1.5$, depending on the age of the stellar bulge of the galaxy~\cite{Bertone:2004pz}. 

Nevertheless, as we discuss next, the differences between the idealized spike and a shallower spike induced by various effects turn out to be often negligible once we consider the regularization due to sizable self-annihilation cross sections. 

We consider a rather conservative scenario where the DM spike has been depleted into a cored-like profile due to DM self-annihilations over sufficiently long timescales. The effect is governed by the following differential equation, tracking the time evolution of DM number density at fixed radius, $\rho_{\rm  DM}  ( r,  t )$:
\begin{align}
  \frac{\dot{\rho}_{\rm  DM}  ( r,  t )}{m_{\rm  DM}}
  &=
  -  \Braket{\sigma_{\rm ann}  v}    \left( \frac{\rho_{\rm DM}  (r, t)}{m_{\rm  DM}} \right)^2,
  \label{eq:DM_number_dissolution}
\end{align}
where $\Braket{\sigma_{\rm ann}  v}$ is velocity weighted DM annihilation cross section. Once we specify the initial DM density for this time evolution, $\rho_{\rm  DM} (r, t_f)$, we obtain the analytic solution, 
\begin{align}
  \rho_{\rm  DM}  (r,t)  
  &=  
  \frac{\rho_{\rm  pl}  (t)  \rho_{\rm  DM}  (r,  t_f)}{\rho_{\rm  pl}  (t)  +  \rho_{\rm  DM}  ( r,  t_f )},
\end{align}
where 
\begin{widetext}
\begin{align} 
 \rho_{\rm  pl}  (t)
   &\equiv  \frac{m_{\rm  DM}}{\Braket{\sigma_{\rm  ann}  v}  ( t  -  t_f )}
   \label{eq:rho_pl}
  \\
  &=  1.1  \times  10^{9}~
  \mathrm{GeV} \mathrm{cm}^{-3}
  \left( \frac{m_{\rm  DM}}{1~\mathrm{GeV}} \right) 
  \left( \frac{ \Braket{\sigma_{\rm  ann}  v} }{3  \times 10^{-26}~\mathrm{cm}^3~\mathrm{s}^{-1}} \right)^{-1}
  \left( \frac{t  -  t_f}{1~\mathrm{Gyr}} \right)^{-1}.
\end{align}
\end{widetext}
This DM profile exhibits a plateau of approximately constant density $\sim  \rho_{\rm  pl}$ in the central region. We identify such plateaus with the region where diffusion may mainly occur.

\begin{figure}[tb]
\begin{center}
\includegraphics[width=0.5\textwidth]{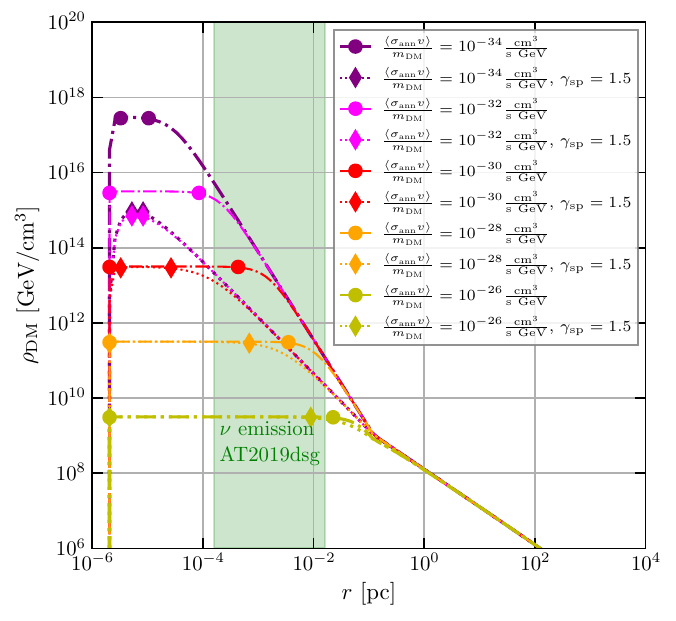}
\end{center}

\caption{DM distribution around a SMBH, including annihilation effects. 
We adopt for the value of $M_{\rm  BH}=5 \times 10^6 M_{\odot}$, consistent with that from the host galaxy of the TDE candidate AT2019dsg. 
Purple, magenta, red, orange, and yellow curves show the DM density with $10^{-34}, 10^{-32}, 10^{-30}, 10^{-28}, 10^{-26}$ for $\Braket{\sigma_{\rm  ann}  v}/m_{\rm  DM}~[\mathrm{cm}^3~\mathrm{s}^{-1}~\mathrm{GeV}^{-1}]$, respectively. 
The dot-dashed (dotted) curves show the spike profile with $\gamma_{\rm  sp}  =  2.33 \cdots (\gamma_{\rm  sp}  =  1.5)$, where the core region is indicated by two circles (diamonds) for each plot. For comparison, we show in shaded green the possible neutrino emission regions for AT2019dsg~\cite{Winter_2023}. }
\label{fig:DM_core}
\end{figure}
%
In Fig.~\ref{fig:DM_core}, we show the DM density distribution for various values of $\Braket{\sigma_{\rm  ann} v}/m_{\rm  DM}$, and taking two spike power exponents ($\gamma_{\rm  sp}  =  2.33  \cdots$, and $1.5$). 
For the black hole mass, we take a fiducial value of $M_{\rm  BH}  \sim  5  \times  10^6~M_\odot$ corresponding to that estimated for the host galaxy of the TDE candidate AT2019dsg \cite{Winter_2023}. A detailed discussion on this and other TDEs is given in Sec.~\ref{sec:application-TDE}, while here we simply take the profiles from Fig. \ref{fig:DM_core} as a benchmark scenario. 
Interestingly, the DM spike's core radii overlaps with potential neutrino emission regions (the green shaded region in Fig.~\ref{fig:DM_core}, taken from \cite{Winter_2023}), as long as $\Braket{\sigma_{\rm  ann}  v}/m_{\rm  DM}  \lesssim  10^{-30}~\mathrm{cm}^3~\mathrm{s}^{-1}~\mathrm{GeV}^{-1}$ depending on the spike exponent. The emission region highly depends on each event and could be closer to the SMBH in some scenarios. 
To determine the normalization of the DM density, we impose that the DM spike mass within the region of gravitational influence (typically $r  \sim  10^5~R_{\rm S}$, where $R_{\rm S}$ is the Schwarzschild radius) is less than the uncertainty of black hole mass ($\Delta  M_{\rm  BH}$) \cite{Gorchtein:2010xa, Lacroix:2015lxa}. In the following analysis, we consider $\Delta  M_{\rm  BH}  \sim  0.1  \times  M_{\rm  BH}$ for each galaxy.
We furthermore take $(t  -  t_f)  =  1~\mathrm{Gyr}$ as a benchmark value for the duration of adiabatic growth of the central black holes in these galaxies.
The central flat DM region, where the plateau density described by Eq.~\eqref{eq:rho_pl} is realized, is specified in the Figure by two circles (diamonds) for $\gamma_{\rm sp}  = 2.33  \cdots$ ($\gamma_{\rm  sp}  =  1.5$). In general, the flat region extends to larger radii for larger annihilation cross sections. Values of the spike's core radius are shown by dashed lines in the left panel of Fig.~\ref{fig:results}, showcasing the importance of the annihilation cross section. 
The core radius depends only on the DM annihilation cross section and are hence vertical. 
The cored region also extends further for steeper spike profiles and for a longer time for $t-t_f$. In such cored regions where the DM density is quasi-constant, neutrinos approximately scatter in an homogeneous DM medium. 

We conclude with discussions on parameter dependencies. 
The SMBH mass differs between galaxies and also impacts the DM density profiles via the spike radius through which the DM density normalization is determined. 
In addition, the inner radius is proportional to $M_{\rm BH}$. 
The smaller the inner radius, the higher the DM density in the inner most region in the case of non-annihilating DM. 
However, for annihilating DM, the plateau DM density crucially determines the highest density, which in turn depends on $\Braket{\sigma_{\rm ann}  v}/m_{\rm DM}$ and $t-t_f$. 
The annihilation cross section is a free parameter in the underlying DM theory. If we consider that the DM solely couples to neutrinos, constraints on the DM self-annihilation cross sections are given by neutrino telescopes sensitive to solar, supernova, atmospheric or high-energy neutrinos~\cite{Arguelles:2019ouk}. 
We will scan over this parameter within the allowed ranges from experiments. If the DM couples to other SM particles besides neutrinos, however, stronger constraints can apply from electromagnetic radiation and cosmic ray telescopes \cite{Slatyer:2021qgc,Cirelli:2023tnx}.
The formation time for the DM density spike $t-t_f$ is not well constrained and is typically estimated as ${\cal  O}  (1-10)~\mathrm{Gyrs}$ \cite{Piana_2020}. 
In the following, we take a benchmark value $t-t_f = 1~\mathrm{Gyr}$. We note that the uncertainty in this parameter is degenerated with the DM annihilation cross section in determining DM density and the size of the plateau region. 
%

\subsection{\label{sec:constraints}Applicability of diffusion regime in DM spikes}

\begin{figure*}[htb]
\begin{center}
\begin{minipage}{1\hsize}
\centering
\begin{minipage}{0.475\hsize}
\includegraphics[width=1\textwidth]{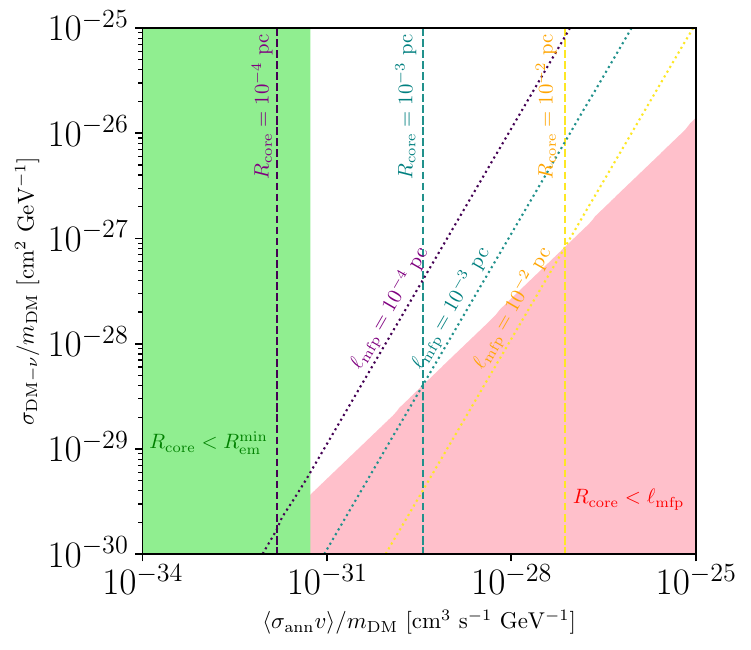}
\end{minipage}
\quad
\begin{minipage}{0.475\hsize}
\includegraphics[width=1\textwidth]{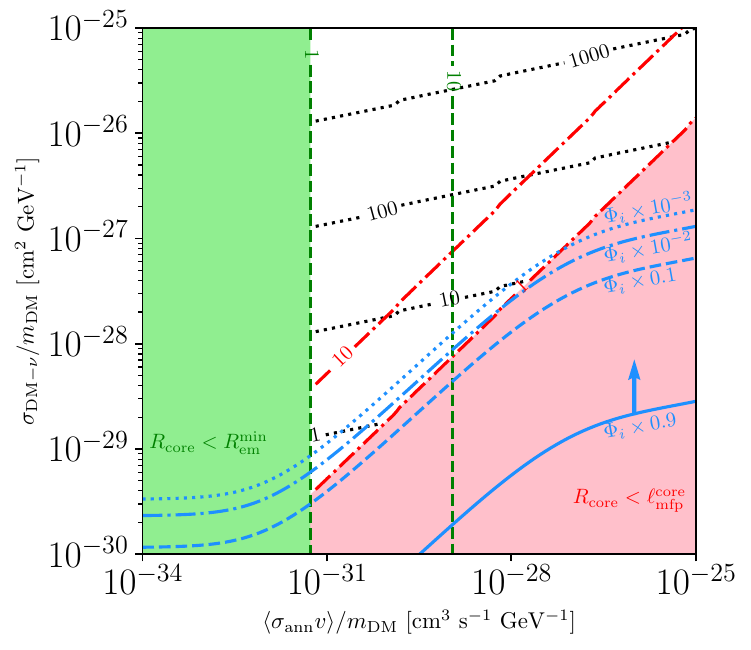}
\end{minipage}
\end{minipage}
\end{center}

\caption{
Cross section dependence of diffusion phenomena, demonstrated using the DM density profile in Fig.~\ref{fig:DM_core} ($\gamma_{\rm  sp}  =  2.33  \cdots$). 
\textit{Left:}
Lines of $R_{\rm  core}$ (dashed) and $\ell_{\rm  mfp}$ (dotted) in units of [pc]. 
In the red-shaded region, the core radius is smaller than the mean free path at the core region. 
The green-shaded region shows where the core radius is smaller than the emission radius for the case of TDE AT2019dsg. Thus, neutrinos may enter the diffusion regime via DM-neutrino scatterings in the non-shaded regions. 
\textit{Right:}
Contours of diffusion time with constraints. 
The black dotted contours show the diffusion time in the unit of [day(s)]. 
The blue contours show the potential attenuation limit via DM-neutrino scattering~\cite{Fujiwara:2023lsv}. 
}
\label{fig:results}
\end{figure*}
%
To realize the diffusive propagation of neutrinos, we consider two conditions that must be met. First, the source of neutrinos must be embedded inside of the high DM density region so that DM-neutrino scattering occurs in the DM spike's core region. That is, $R_{\rm  em} < R_{\rm  core}$, where $R_{\rm  em}$ is the radial location of neutrino source, and $R_{\rm  core}$ is the DM spike's core size. The DM spike's core size depends on the annihilation cross section, as shown in Fig.~\ref{fig:DM_core}, while the radial location of neutrino source varies by source but can be within the DM spike's cores as shown for AT2019dsg in Fig.~\ref{fig:DM_core}. Other potential sources, further studied in Sec.~\ref{sec:applications}, include high-energy activity associated with the SMBH as well as supernovae. 

Second, the mean free path for DM-neutrino scatterings in the DM core ($\ell_{\rm  mfp}$) should be smaller than the core radius, that is, $\ell_{\rm  mfp} < R_{\rm  core}$.
Since the DM density is quasi-constant in the DM core region, the mean free path in that region is also constant as defined,
\begin{align}
  \ell_{\rm  mfp}  =  \frac{m_{\rm  DM}}{\rho_{\rm DM}(R_{\rm  core})  \sigma_{\rm  DM-\nu}}.
  \label{eq:lmfpCore}
\end{align}
We show contours of the mean free path [pc] in the left panel of Fig.~\ref{fig:results}  (dotted lines), compared to the DM spike's core size $R_{\rm  core} \, \mathrm{[pc]}$ (dashed lines). 
The intersection of lines with the same colors corresponds to the boundary of the diffusion condition, which indicates $\ell_{\rm  mfp}  =  R_{\rm  core}$. 
For a fixed scattering cross section, the mean free path gets smaller for smaller annihilation cross section, because a larger DM density is realized in the DM spike's core. 

Note that we neglect DM-neutrino scatterings that occur beyond the DM spike's core, which we typically expect to induce sub-leading corrections. This is due to the fact that, while the distance scale is increased, the DM density decreases faster due to the steep index, \textit{e.g.,} $\gamma_{\rm  sp} \gtrsim 2$ in the spike. 
Similarly, we conservatively neglect the cases where the point of neutrino emission is on the opposite side of SMBH from the observer.
In this case, neutrino should propagate avoiding the SMBH to reach to the observer's point, but given the larger distance through the DM spike's core, it will lead to more scatterings. 

\subsection{Effects of neutrino diffusion}\label{sec:delay}

Based on the DM spike's density and condition for neutrino diffusion, we explore three effects on the neutrino signature: temporal, intensity, and energy. In Sec.~\ref{sec:applications}, we will critically apply what we find to specific sources, in particular TDEs where there exists an observed hint of delay in neutrino arrival times.

\subsubsection{Temporal effect: delay and spread in arrival}

As the neutrinos scatter randomly multiple times on their way out of the DM spike's core, their evolution can be modeled according to the diffusion equation. 
Assuming isotropic scattering, which we justify given the quasi-uniform distribution of the target DM density, we can characterize the diffusion time as 
\begin{align}
  \tau  \sim  \frac{3  R_{\rm  sys}^2}{2  c}  \times  n_i  \times  \sigma_{i-\nu},
  \label{eq:diff_time}
\end{align}
where $n_i$ and $\sigma_{i-\nu}$ are the number density of target ($i$) and its scattering cross section with a neutrino, respectively. We consider the main contribution to scattering to occur in the DM spike's core region, where the highest DM density is realized. Then, we can estimate the neutrino's time delay based on a simple application of Eq.~\eqref{eq:diff_time} to the DM spike's core,
\begin{align}
  \tau_{\rm  DM-\nu}
  &=
  \frac{3  R_{\rm  core}^2  }{2  c}
  \times  
  \rho_{\rm  DM}(R_{\rm  core})
  \times  
  \frac{\sigma_{\rm  DM-\nu}}{m_{\rm DM}}, 
      \label{eq:time_delay-DM}
\end{align}
where the value of DM density and the system size are evaluated as the quasi-constant DM density (Eq.~\eqref{eq:rho_pl}) and its size  ($\sim  R_{\rm  core}$), respectively. 

Depending on the steepness of the spike, taking larger regions and averaging can be more beneficial than taking smaller regions without averaging. This occurs in particular in scenarios where the DM spike has steepness $\gamma_{\rm sp} \leq 2$, since the diffusion timescale scales with the spike's core size as $\tau \sim R_{\rm core}^2$. 
On the other hand, when the density spike decreases with steeper index $\gamma_{\rm  sp} \gtrsim 2$, we expect the change to the time delay to be sub-leading.   
We fix the original spike power law $\gamma_{\rm  sp}  =  2.33  \cdots$ to avoid the uncertainty in estimation of $R_{\rm  sys}$, and adopt the radius of the DM spike's core as the system size by focusing on the case where the neutrino emission radius is smaller than or comparable to the DM spike core radius. 
This gives a reasonably conservative estimate of the neutrino traveling distance through the DM. 

In the right panel of Fig.~\ref{fig:results}, we show the parameter region where DM-neutrino scattering enters the diffusion regime at the DM core and the estimated diffusion time. 
We use the same DM density with $\gamma_{\rm  sp}  =  2.33  \cdots$ shown in Fig.~\ref{fig:DM_core}. 
Here, we assume that the energy dependence in DM-neutrino scattering cross section can be negligible to avoid energy reduction via multiple scatterings (however, see Sec.~\ref{sec:application-TDE}). 
We shade the regions where the diffusion regime cannot be achieved at the DM core. The green (red) region are defined by $R_{\rm  core}/R_{\rm  em}$ ($R_{\rm  core}/\ell_{\rm  mfp}$). 
For $R_{\rm  em}$, we take the minimum value of emission radius for AT2019dsg within its uncertain range, and the green contours/region depend on each neutrino source. 
We apply Eq.~\eqref{eq:time_delay-DM} to estimate the delay in between green and red shaded regions. 
The black dotted contours show the diffusion time $\tau_{\rm  DM-\nu} \, \mathrm{[days]}$.
In general, longer times are realized for larger scattering cross sections, but also for smaller annihilation cross sections because of its impact on the DM spike's core density. 

\subsubsection{Flux attenuation}

If the DM-neutrino scattering cross section is sufficiently large, the reduction of the emitted neutrino flux during propagation through the DM spike or core can be significant, as explored by Refs.~\cite{Ferrer:2022kei, Cline:2022qld, Fujiwara:2023lsv, Cline:2023tkp}. 
The injected flux and the flux with DM-neutrino scattering can be related by solving the cascade equation~\cite{Vincent:2017svp,Arguelles:2017atb,Cline:2022qld, Fujiwara:2023lsv},
\begin{align}
  \frac{d  \Phi}{d  \tau}  =  -  \sigma_{\rm DM-\nu}  \Phi  +  \int_{E_\nu}^\infty  d  E_\nu'  \frac{d  \sigma_{\rm DM-\nu}}{d  E_\nu}  ( E_\nu'  \to  E_\nu )  \Phi  (E_\nu').
  \label{eq:cascade_eq}
\end{align}
where we define $\tau  =  \Sigma_{\rm DM}/m_{\rm  DM}$, and $\Sigma_{\rm DM}$ denotes the DM column density along the line of sight, 
\begin{align}
  \Sigma_{\rm  DM}  =  \int_{\rm  l.o.s}  dr  \rho_{\rm  DM}  (r).
  \label{eq:column_density}
\end{align}
By considering the theoretically expected high-energy neutrino fluxes from astrophysical sources within their associated uncertainties, and the observed fluxes at Earth-based detectors like IceCube, one can place a constraint on the largest allowed DM-neutrino scattering cross section.
In particular, regarding TDEs, efforts have been made to compare model predictions with the IceCube hint~\cite{Winter_2023}. 
Based on these models, the DM-neutrino scattering cross section can be potentially constrained from TDEs~\cite{Fujiwara:2023lsv}, although the uncertainties on the initially injected neutrino fluxes at the sources are still large depending on astrophysical models.

In the right panel of Fig.~\ref{fig:results}, we show the impact of the flux attenuation. 
For a quasi-constant scattering cross section, we can neglect the second term of Eq.~\eqref{eq:cascade_eq}, and the suppression of the emitted fluxes is controlled by an exponential factor. 
If we require the initial flux should not be reduced more than 10\%, 
the scattering cross section should satisfy 
\begin{align}
  \Sigma_{\rm  DM}  \frac{\sigma_{\rm  DM-\nu}}{m_{\rm  DM}}  \lesssim  0.1.
  \label{eq:attenuation}
\end{align}
This inequality is satisfied for the region above the blue solid curve. 
The initial flux will be reduced to $10^{-1}, 10^{-2}$, and $10^{-3}$ for $\sigma_{\rm  DM-\nu}/m_{\rm  DM}  \lesssim  2.3, 4.6$, and $6.6$, which is indicated in the blue dashed, dotdashed, and dotted curves, respectively.

Notice that the diffusion region (above the red-shaded area) and the region for sizable attenuation (above the blue dashed curve) are well-aligned. 
This correlation can be clarified by noting the column density can be dominated by the DM core region, where we have the highest DM density.
If we focus on the core region and use Eq.~\eqref{eq:column_density}, we obtain 
\begin{align}
  \left.
  \int_{\rm  l.o.s}  dr  \rho_{\rm  DM}  (r)  \frac{\sigma_{\rm  DM-\nu}}{m_{\rm  DM}}
  \right|_{\rm  core}
  \sim  
  R_{\rm  core}  
  \cdot
  \frac{1}{\ell_{\rm  mfp}^{\rm core}},
  \label{eq:attenuation_vs_diffusion}
\end{align}
where we use 
Eq.~\eqref{eq:lmfpCore}. 
Therefore, we can restate the condition for the initial flux to be attenuated more than one order of magnitude as $R_{\rm core}  \gtrsim  {\cal  O}  (1)  \times  \ell_{\rm  mfp}$, which strongly correlate the diffusion parameter space where $R_{\rm core}  >  \ell_{\rm  mfp}$.
We find that this corelation is broken for $\Braket{\sigma_{\rm   anm}  v}  \gtrsim  10^{-27}~\mathrm{cm^3/s}$ and $\Braket{\sigma_{\rm   anm}  v}  \lesssim  10^{-31}~\mathrm{cm^3/s}$ in Fig.~\ref{fig:results}. 
For the former region, the core density has decreased enough such that the core region is less dominant than the rest of the halo to the total column density. 
For the latter region, the smaller core radius makes the core contribution less relevant. 
We stress that Eq.~\eqref{eq:attenuation_vs_diffusion} holds as long as the column density is dominated by the core region, regardless of parameters in $\rho_{\rm  DM}$ such as $M_{\rm BH}$ and $\gamma_{\rm   sp}$. 
Therefore, the correlation between the diffusion parameter space and the attenuation region is a quite general feature. 
This implies that the DM contribution to the neutrino diffusive delay always associates the significant flux attenuation. 

\begin{figure}[tb]
\centering
\includegraphics[width=0.5\textwidth]{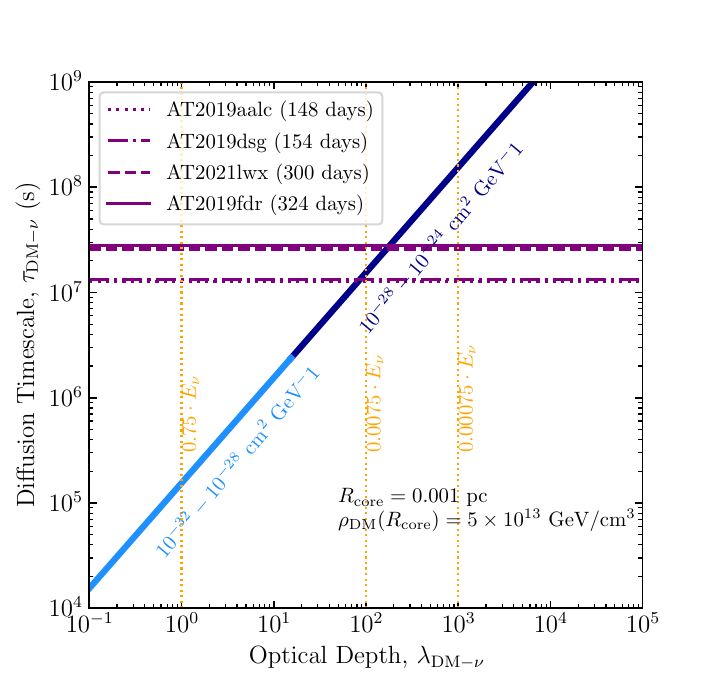}
\caption{Diffusion timescale confronted by DM-neutrino scattering optical depth in a typical DM spike's core with size $R_{\rm core}=10^{-4}$ pc and DM density $\rho_{\rm DM}=10^{16}$GeV/cm$^3$. We show contours corresponding to different ranges of $\sigma_{\rm DM-\nu}/m_{\rm DM}$, and confront them with the suggested neutrino delays from various TDEs. Furthermore, we show vertical orange dotted lines indicating the average energy of neutrinos after the relvant number of scatterings for a given value of the optical depth, relative to the initial energy $E_{\nu}$.}
\label{fig:delays_vs_attenuation}
\end{figure}

\subsubsection{Energy degradation}\label{sec:energy}

In each DM-neutrino scattering, some of the neutrino energy is imparted onto the DM, which is effectively at rest. Furthermore, if we consider a realistic model of DM-neutrino scattering, the cross section can strongly depend on the neutrino energy $E_\nu$. These aspects lead to changes in the neutrino signal's energy distribution.

The neutrino energy can be rapidly transferred to the DM via scattering, depending on the scattering angles. 
In our system, we can take the rest frame of the DM with an incoming flux of neutrinos. We obtain the following kinetic energies for the neutrino and DM after scattering, 
\begin{align}
T_\nu  &=  E_{\rm  CM}  -  m_{\rm  DM} - T_{\rm  DM},
\\
T_{\rm  DM}  &=  \frac{(T_\nu^{\rm  ini})^2}{T_\nu^{\rm  ini}  +  m_{\rm  DM}/2}  \times  \frac{1  -  \cos  \theta_\star}{2},
\end{align}
where $E_{\rm CM}$, $T_\nu^{\rm  ini}$, and $\theta_\star$ are the center mass energy, initial neutrino kinetic energy, and scattering angle at the center mass frame, respectively. 
We neglect the neutrino mass in this formula. 
For $T_\nu^{\rm  ini}  \gg  m_{\rm  DM}$, the transferred energy is of the order of the initial neutrino energy (if we take average of the angle, we find $T_\nu/4$). 
In the diffusion regime, neutrinos scatter multiple times with the DM and may lose significant energy. This effect is shown in Fig.~\ref{fig:delays_vs_attenuation} as vertical dotted orange lines. For representative values of the optical depth, we show the corresponding average energy of the neutrinos after undergoing the relevant number of scatterings. It can be appreciated that for sufficiently large values of the optical depth, the average energy lost by neutrinos can be of orders of magnitude. For emitted neutrino fluxes that follow a falling power-law with increasing energy, such an effect can prevent the final neutrino fluxes to peak at high energies and may be constrained by observations at, \textit{e.g.,} IceCube.

\section{\label{sec:applications}
Application to specific neutrino sources}

Interestingly, we find that DM-neutrino scatterings can induce a significant time delay and spread in the neutrino signal, at a level similar with the observed hints of delays in TDEs ~\cite{Winter_2023}. However, this would come at the cost of attenuation in the flux and reduction in neutrino energies. 

In order to allow for comparison of the expected neutrino delays with complementary probes on neutrino DM-scatterings, we address the TDE time delay from DM-neutrino scatterings in a concrete DM model. 
Indeed, we critically find that observable (and large) delays at Earth can still be induced by unconstrained new physics.

\subsection{\label{sec:application-TDE}
Tidal Disruption Events}

\begin{table}[tb]
  \renewcommand{\arraystretch}{1.5}
  \begin{center}
  \begin{tabular}{c||c|ccc}
  \hline
  TDE
  &  $M_{\rm BH}~[M_{\odot}]$ 
  &  $R_{\rm em}~[\mathrm{pc}]$
  &  $\tau_{\rm  delay}~[\mathrm{days}]$ 
  \\
  \hline
  AT2019dsg
  &  $5  \times  10^{6}  M_{\odot}$ 
  &  $1.6 \times 10^{-4}$ -- $0.016$
  &  154
  \\
  AT2019fdr
  &  $1.3  \times  10^{7}  M_{\odot}$ 
  &  $0.0015$ -- $0.81$
  &  324
  \\
  AT2019aalc
  &  $1.6  \times  10^{7}  M_{\odot}$ 
  &  $0.0016$ -- $0.065$
  &  148
  \\
  AT2021lwx
  &  $1  \times  10^{8}  M_{\odot}$ 
  &  $0.17$ -- $0.32$
  &  370
                  \\
                \hline
			\end{tabular}
		\end{center}
		\caption{Relevant parameters for the TDE candidates. 
		$M_{\rm BH}$ is the mass of the central black hole, 
		$R_{\rm em}$ is the expected emission radii for neutrino~\cite{Winter_2023}, 
		and 
		$\tau_{\rm delay}$ is neutrino delay compared to the peak time of photon luminosity.
		The values of $\tau_{\rm delay}$ are taken from Table~1 in Ref.~\cite{Winter_2023}.
		}
		\label{tab:TDEs}
	\end{table}

TDEs occur when an orbit of a massive star gets close to the SMBH. 
The tidal force of the black hole disrupts the star, and its debris falls back to the SMBH, which shows a characteristic flare for months to years. The tidal disruption radius is typically of order ten to hundred times the Schwarzschild radius, thus lying within the expected extension of the DM spike. 
Currently, four neutrino events at IceCube (IC191001A, IC200530A, IC191119A, IC220405B) are associated with the optical counterparts identified by the Zwicky Transient Facility (AT2019dsg, AT2019fdr, AT2019aalc, AT2021lwx, respectively). Interestingly, these four events show neutrino delays around ${\cal  O}(100)~\mathrm{days}$ with respect to the blackbody photon peak time. Some complex astrophysical models can explain the delay in certain circumstances, see, \textit{e.g.,} Ref.~\cite{Winter_2023}.
We will apply the general discussion above to TDEs and estimate time delay induced by DM-neutrino scatterings. 
The parameters used to fix the DM density profile are shown in Table~\ref{tab:TDEs}. For a detailed explanation on the DM distribution around the SMBH of TDEs (AT2019dsg, AT2019fdr, and AT2019aalc), we refer the reader to Ref.~\cite{Fujiwara:2023lsv}.

Our previous estimates have been performed assuming that the DM-neutrino scattering cross section is constant at all relevant neutrino energies. 

It is worth considering instead a concrete model of a Dirac fermion DM particle, and a vector mediator with mass $m_{Z'}$, which in general exhibits an energy dependence in the scattering cross section. 
Specifically, the DM-neutrino scattering cross section in terms of neutrino energy $E_{\nu}$ in this model reads \cite{Fujiwara:2023lsv}:
\begin{widetext}
\begin{align}\label{eq:cross_section_vector}
    \sigma_{\rm DM\text{-}\nu} 
    &
    = \frac{g_{\nu}^2 g_{\rm DM}^2}{16 \pi E_\nu^2 m_{\rm DM}^2}
    \Bigl[ 
    (2 E_\nu m_{\rm DM} + m_{{Z^{\prime}}}^2 + m_{\rm DM}^2) \log \left( \frac{m_{Z^{\prime}}^2 (2 E_\nu + m_{\rm DM})}{4 E_\nu^2 m_{\rm DM} + 2 E_\nu m_{{Z^{\prime}}}^2 + m_{\rm DM} m_{{Z^{\prime}}}^2} \right) 
    \nonumber
    \\
    &~~~~~~~~~~~~~~~~~~~~~~
    + 4 E_\nu^2 \left( 1 + \frac{m_{\rm DM}^2}{m_{{Z^{\prime}}}^2} - \frac{2 E_\nu (4 E_\nu^2 m_{\rm DM} + E_\nu (m_{\rm DM}^2 + 2 m_{{Z^{\prime}}}^2) + m_{\rm DM} m_{{Z^{\prime}}}^2)}{(2 E_\nu + m_{\rm DM}) (m_{\rm DM} (4 E_\nu^2 + m_{{Z^{\prime}}}^2) + 2 E_\nu m_{{Z^{\prime}}}^2)} \right) \Bigr],
\end{align}
 \end{widetext}
where $g_{\nu}$, $g_{\rm DM}$ represent the couplings of the new mediator to the neutrinos and to the DM, respectively.
This model-dependent analysis allows us to compare diffusion phenomena at TDEs with other complementary constraints on light neutrinophilic mediators.

In Fig.~\ref{fig:delays_vs_attenuation}, we confront the diffusion timescale vs optical depth from DM-neutrino scatterings in a typical DM core from these neutrino-emitting TDEs, for different values of the DM-neutrino scattering cross section, where the optical depth in the core region is defined as
\begin{equation}
\lambda_{\rm DM-\nu}=\frac{\rho_{\rm DM}}{m_{\rm DM}}(R_{\rm core})\sigma_{\rm DM-\nu} R_{\rm core}.
\end{equation}
It can be appreciated that entering the diffusion regime yields delay timescales larger than $\tau_{\rm DM-\nu} \gtrsim 10^5$ s. However, if aiming to yield a delay timescale compatible with the hints observed from the TDEs AT2019aalc, A72019dsg, AT2019fdr and AT2021lwx, the corresponding optical depth is larger than $\gtrsim 60$, which would cause an attenuation of the initially emitted neutrino fluxes of several orders of magnitude, thus unlikely. This plot indicates that the suggested neutrino delays from TDEs cannot be fully explained via DM-neutrino scatterings at the source. However, the diffusion timescale can be sizable, of order ${\cal O} (0.1-10)$ days, for reasonable optical depths of ${\cal O} (1-10)$.

\begin{figure}[tb]
\centering
\includegraphics[width=0.5\textwidth]{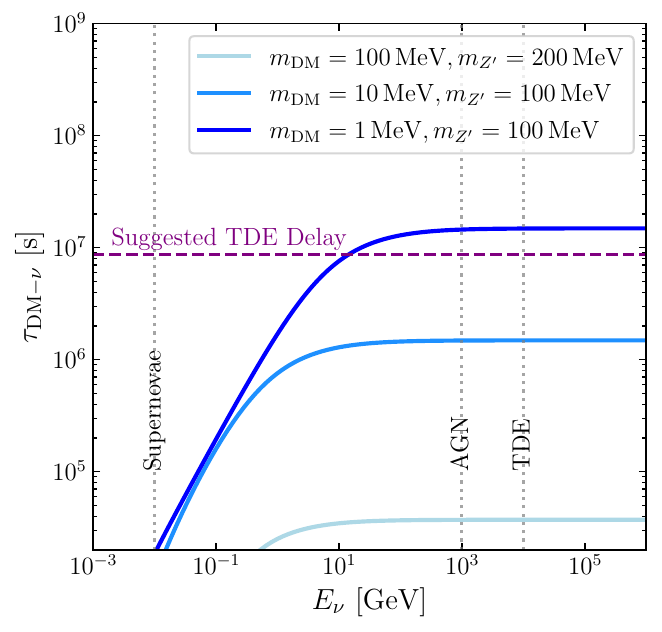}
\caption{Diffusion timescale induced by DM-neutrino interactions via a vector mediator confronted with neutrino energy, for a typical spike with size $R_{\rm core}=10^{-4}$ pc, and with DM density $\rho_{\rm DM}=10^{16}$GeV/cm$^3$. The diffusion condition is satisfied for all the lines shown in the plot. We fix the products of gauge couplings of the mediator to neutrinos and DM as $g_{\rm DM}g_{\nu}=0.01$. We confront the diffusion timescales induced by various values of the mediator mass with the tentative delays observed from the TDEs considered in this work.}
\label{fig:energy_dependence}
\end{figure}

We show this dependence in Fig.~\ref{fig:energy_dependence} for a model with a Dirac fermion DM particle and a vector $Z^{\prime}$ mediator (see Eq. \eqref{eq:cross_section_vector} ), 
for different values of the DM mass and mediator mass. We fix the coupling strength of the scattering to $g_{\nu}g_{\rm DM}=0.01$. For comparison, we show suggested delays from TDEs. In the plot it can be clearly appreciated that at low energies the delay may be shorter, while constant at sufficiently large energies. If the emitted neutrino spectra follows a falling power-law spectra, it is likely that the delay in neutrino arrival would only be seen above a certain neutrino energy threshold, while the low-energy part of the neutrino emission arrives as expected.
We also note that the diffusion condition at the core ($\ell_{\rm  mfp}  <  R_{\rm  core}$) should be carefully examined as $\ell_{\rm 
 mfp}$ depends on $E_\nu$ during scattering via energy dependence of cross section.

\begin{figure}[tb]
\centering
\vspace{1.5mm}

\includegraphics[width=0.5\textwidth]{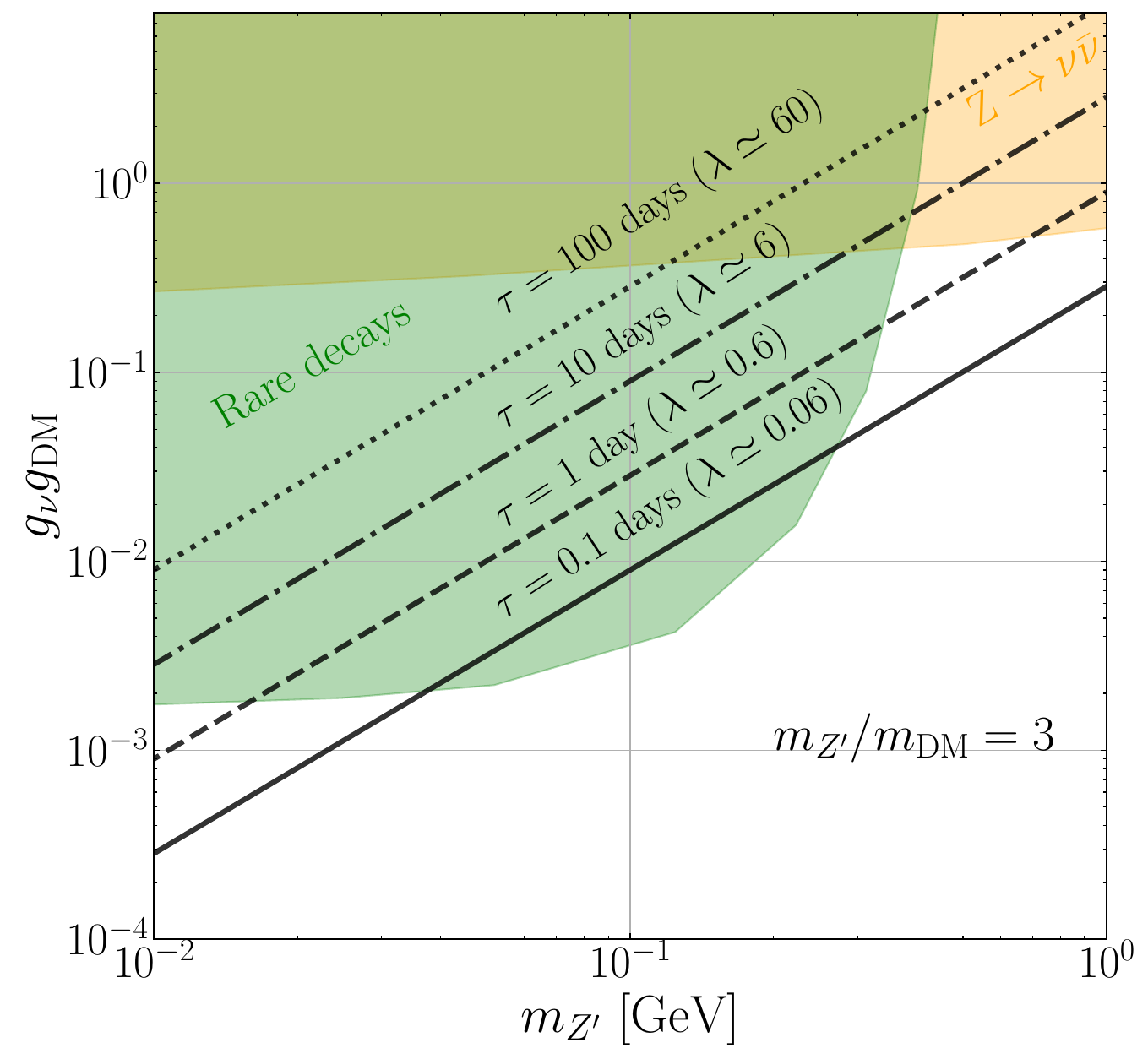}
\caption{Expected diffusion times induced by DM-neutrino scatterings in AT2019dsg, for $\left\langle\sigma_{\mathrm{ann}} v\right\rangle / m_{\mathrm{DM}}=10^{-30} \mathrm{~cm}^3 \mathrm{~s}^{-1} \mathrm{GeV}^{-1}$ and $R_{\rm core}=0.001$ pc. The results are displayed in the parameter space spanned by the couplings of the new mediator to the DM and neutrino sectors vs mediator mass, for a given ratio of DM and mediator masses. These parameters can be related to the usual non-relativistic constant scattering cross section as $\sigma_{\rm DM-\nu} \simeq g_{\nu}^2g_{\rm DM}^2m_{\rm DM}^2/4\pi m_{Z'}^4$. We confront our results with bounds from Z boson decays into neutrinos, and rare meson decays \cite{Berryman:2022hds,Dev:2024twk}.}
\label{fig:delays_mediator}
\end{figure}

Finally, it is worth translating these estimates on the expected neutrino delays from diffusion within DM spikes and cores into the parameter space spanned by the product of couplings and the mediator mass, to allow for comparison with complementary probes of neutrinophilic mediators, see, {\rm \textit{e.g.}}~\cite{Berryman:2022hds,Dev:2024twk}. In Fig.~\ref{fig:delays_mediator}, we compare current constraints from rare meson decays and $Z$ boson decays into neutrinos with the expected neutrino delays ($\tau$) and optical depth ($\lambda$) for different choices of product of couplings and mediator mass. We fix $m_{Z^{'}}/m_{\rm DM}=3$. We have demonstrated in this plot that, for a concrete model of DM-neutrino interactions, there remain regions of parameter space where sizable delays on neutrino arrival from astrophysical sources can be induced via diffusion in a DM medium, while respecting complementary constraints. Correspondingly, future timing information from astrophysical neutrino sources can allow to constrain new regions of parameter space untested by laboratory probes.

\subsection{\label{sec:future_events}
Future events}
Future observations may indicate delays from sources other than TDEs. Neutrino emitting transients like the blazar TXS 05056+056 already featured a mismatch on the neutrino and electromagnetic neutrino arrival in two flares \cite{IceCube:2018cha}. The neutrino signal from TXS 0506+056 did not feature a significant delay, unlike TDEs, but this may be caused by the neutrino alert being identified first, and electromagnetic telescopes took some time to point in the direction of TXS 0506+056. The observation of future blazars in both high-energy neutrino and electromagnetic emission could feature astrophysically inexplicable delays, and our discussion may be pertinent. Future correlation searches from IceCube of blazar and AGN catalogs will be useful in this regard \cite{IceCube:2023htm, Bellenghi:2023yza}. Other transients like Gamma-Ray Bursts could also induce neutrino delays \cite{Kimura:2022zyg, Murase:2022vqf}. A future galactic supernovae could also provide useful information \cite{Adams:2013ana, Horiuchi:2018ofe, Wen:2023ijf}. Of particular interest would be Galactic transients occurring near the vicinity of Sagittarius A$^{*}$, where the DM density is likely to be enhanced.

\section{\label{sec:conclusions}
Conclusions and discussions}

We have estimated for the first time the diffusion time induced by DM-neutrino scatterings in astrophysical sources embedded in a DM spike's core. We have shown that the diffusion time can be large, yielding significant delays and spreading of the neutrino signal observed at Earth. In particular, we demonstrated that for astrophysical transients that have been previously studied in the context of significant suppression of the emitted neutrino flux, the time delays can be of order $\sim$ days. We explored these effects both for a constant DM-neutrino scattering cross section and for an energy-dependent cross section in a concrete model with a vector mediator, showing that significant delays can be induced at the source in regions of the parameter space untested by complementary probes on neutrinophilic mediators.

We focused our discussion on neutrino-emitting TDEs, where a hint of time delay of the neutrino arrival with respect to the peak of the electromagnetic emission of $\sim 100$ days has been observed. Such delays are unlikely to be fully caused by DM-neutrino scatterings, since it would imply an attenuation of the emitted astrophysical neutrino fluxes of several orders of magnitude. However, we cannot preclude that a fraction of the total delay may be induced by DM-neutrino scatterings, since there are uncharted regions of parameter space of DM models where this is possible.

The multimessenger era is bringing exciting high-energy neutrino data that open windows for new physics. The timing information from both neutrinos and photons, coupled with the neutrino fluxes and energy spectra, allows to not only better understand the underlying physics at the source, but potentially revealing hints for new physics yet obscured to us. We hope that our work contributes to a better understanding of the neutrino delays expected from DM at the sources themselves.

\section*{Acknowledgments}
\noindent
The work of MF was supported by the Collaborative Research Center SFB1258 and by the Deutsche Forschungsgemeinschaft (DFG, German Research Foundation) under Germany’s Excellence Strategy-EXC-2094-390783311.
The work of GH and SH is supported by the U.S. Department of Energy Office of Science under award number DE-SC0020262. 
The work of SH is also supported by NSF Grant No. AST1908960 and No.~PHY-2209420, and JSPS KAKENHI Grant Number JP22K03630 and JP23H04899.

\bibliography{references}

\end{document}